\documentclass[12pt,preprint]{aastex}
\usepackage{emulateapj5} 

\newcommand{\kms}{$\,\mbox{km}\,\mbox{s}^{-1}$}

\newcommand{\etal}{{et al.\ }}
\newcommand{\Msun}{M$_{\odot}$}

\shortauthors{Walter, Martin \& Ott}
\shorttitle{Star Formation in the Tidal Arms near NGC\,3077}

\begin{document}

\title{Extended Star Formation and Molecular Gas in the Tidal Arms
near NGC\,3077}

\author{Fabian Walter} 
\affil{Max Planck Institut f\"ur Astronomie}
\affil{K\"onigstuhl 17, 69117 Heidelberg, Germany}
\email{walter@mpia.de}
\vspace{0.5cm}
\author{Crystal L. Martin$^{1,2}$} 
\affil{University of California, Santa Barbara}
\affil{Santa Barbara, CA 93106, USA}
\email{cmartin@physics.ucsb.edu}
\altaffiltext{1}{Packard Fellow}
\altaffiltext{2}{Alfred P. Sloan Research Fellow}

\and

\author{J\"urgen Ott$^3$} 
\affil{Australia Telescope National Facility}
\affil{Cnr Vimiera \& Pembroke Roads, Marsfield NSW 2122, Australia}
\email{Juergen.Ott@csiro.au}
\altaffiltext{3}{Bolton Fellow}

\begin{abstract}
  
  We report the detection of ongoing star formation in the prominent
  tidal arms near NGC\,3077 (member of the M\,81 triplet).  In total,
  36 faint compact \ion{H}{2} regions were identified, covering an
  area of $\sim4\times6$ kpc$^2$. Most of the \ion{H}{2} regions are
  found at \ion{H}{1} column densities above
  $1\times10^{21}$\,cm$^{-2}$ (on scales of 200\,pc), well within the
  range of threshold columns measured in normal galaxies. The
  \ion{H}{2} luminosity function resembles the ones derived for other
  low--mass dwarf galaxies in the same group; we derive a total star
  formation rate of 2.6$\times10^{-3}$ \Msun\,yr$^{-1}$ in the tidal
  feature. We also present new high--resolution imaging of the
  molecular gas distribution in the tidal arm using CO observations
  obtained with the OVRO interferometer. We recover about one sixth of
  the CO flux (or M$_{H2}\sim2\times10^6$\,\Msun, assuming a Galactic
  conversion factor) originally detected in the IRAM 30\,m single dish
  observations, indicating the presence of a diffuse molecular gas
  component in the tidal arm. The brightest CO peak in the
  interferometer map (comprising half of the detected CO flux) is
  coincident with one of the brightest \ion{H}{2} regions in the
  feature. Assuming a constant star formation rate since the creation
  of the tidal feature (presumably $\sim3\times10^8$\,years ago), a
  total mass of $\sim7\times10^5$ \Msun\ has been transformed from gas
  into stars. Over this period, the star formation in the tidal arm
  has resulted in an additional enrichment of $\Delta$Z$>$0.002. The
  reservoir of atomic and molecular gas in the tidal arm is
  $\sim3\times10^8$ \Msun, allowing star formation to continue at its
  present rate for a Hubble time.  Such wide--spread, low--level star
  formation would be difficult to image around more distant galaxies
  but may be detectable through intervening absorption in quasar spectra.
\end{abstract}


\section{Introduction}

Studies of the processes that lead to star formation are critical for
our understanding of how galaxies form and evolve with time. In
massive galaxies, star formation is triggered through a variety of
processes: e.g., compression of molecular cloud cores in spiral
density waves, tidal interactions, and bar formation. On scales of a
few hundred parsecs, star formation is observed where the gas density
exceeds a threshold, whose value varies with the critical density for
local gravitational stability (e.g., Martin \& Kennicutt 2001). Models
that rely on thermal instabilities to determine the star formation
threshold predict a constant threshold column density (Schaye 2004).

Studies of star formation (SF) in tidal arms of interacting galaxies
offer the possibility to study SF processes at conditions that are
distinctly different from those in normal galactic disks (e.g., given
the absence of spiral density waves and bars).  Nearby, interacting
systems give us the unique opportunity to study the properties of
tidal features (e.g.  atomic and molecular gas content and \ion{H}{2}
region properties) in detail.  Tidal dwarf candidates have been
detected in tidal tails of interacting galaxies in a number of cases
(e.g., see list in Braine et al.\ 2001); but selection effects limit
these to bright, and therefore fairly evolved, tidal dwarfs. The
detection of SF regions around interacting systems (`intergalactic'
\ion{H}{2} regions) has been reported in Ryan--Weber et al.\ (2004).
Given the much higher incidence of interacting galaxies in the past, a
significant fraction of today's dwarf galaxies may have formed out of
tidal debris (Hunsberger \etal 1996, Hunter \etal 2000, Okazaki \&
Taniguchi 2000, Duc et al.\ 2004).  Also of interest in this context
is the self-enrichment of these tidal debris, and their possible
incidence as intervening metal line systems.

The nearby M\,81 triplet (consisting of M\,82, M\,81 and NGC\,3077) is
an ideal testbed to study tidal arms around interacting galaxies in
detail.  The \ion{H}{1} map of Yun et al. (1994) shows a geometrical
cross section of roughly 50 kpc $\times$ 100~kpc at column densities of
N(HI)$>$1.8$\times$10$^{20}$ cm$^{-2}$.  The impressive tidal arm near
NGC\,3077 is the subject of the present study (total \ion{H}{1} mass:
$\sim5\times10^8$\,\Msun, Walter et al.\ 2002).  According to various
numerical simulations (Thomasson \& Donner 1993, Brouillet \etal 1991,
Yun \etal 1993) this tail has been created by a recent
($\sim3\times10^8$\,yr) interaction with M\,81.  Barbieri et al.
(1974) detected `a fragmentary complex of almost stellar objects' in
this region (the `Garland', see also Karachentsev \etal 1985, Sharina
1991, Sakai \& Madore 2001). The presence of extended molecular gas in
this tidal feature has been reported by Walter \& Heithausen (1999)
and Heithausen \& Walter (2000). In this paper we report on the
discovery of low--level star formation over roughly 24~kpc$^2$ and
present new high--resolution observations of the molecular gas
distribution in this feature.  In Section 2 we summarize our
observations obtained at the 2\,m telescope at Kitt Peak, and at the
OVRO millimeter interferometer. In Sec.~3 we present the distribution
of the \ion{H}{2} regions, the \ion{H}{2} luminosity function and the
high--resolution CO maps; Sec.~4 summarizes our results.

\section{Observations}

\subsection{Optical Observations}

We obtained deep narrowband imaging of the region including NGC\,3077
and the prominent tidal arm with the 2\,m telescope at Kitt Peak on
2000 January 2--10. The seeing yielded a median image quality of
1\farcs3 FWHM. The individual frames were corrected for fixed pattern
noise in the standard way using IRAF\footnote{IRAF is distributed by
  the National Optical Astronomy Observatory, which is operated by the
  Association of Universities for Research in Astronomy, Inc., under
  cooperative agreement with the National Science Foundation.},
aligned using foreground stars, and combined by filter subset.  The
H$\alpha$ image presented here was constructed from an on-band image
taken through filter KP1468, central wavelength (CWL) of $\lambda
6567$ and a full width half maximum (FWHM) of 87\AA, and an
appropriately-scaled continuum image taken through filter KP809 (CWL =
6488\AA, FWHM=67\AA).  Observations of spectrophotometric standards
were used to flux calibrate the H$\alpha$ image. The H$\alpha$ image
contains [NII]~$\lambda \lambda 6548, 6584$ emission.  The correction
is H$\alpha$/([NII]+H$\alpha$)=0.75 in the center of NGC\,3077 and
increases to H$\alpha$/([NII]+H$\alpha$) = 0.56 in the extended gas,
where the ionization parameter is low (Martin 1997).  This correction
has not been measured in the tidal gas; but if the gas phase
metallicity is lower in the arm, then the contribution from [NII] will
be smaller than in NGC\,3077. The foreground Galactic extinction is
about 0.11 magnitudes at H$\alpha$ (Schlegel et al. 1998; Burstein \&
Heiles 1982).

\subsection{OVRO observations}

We observed NGC\,3077 in the CO(1$\to$0) transition at 115\,GHz
($\sim$2.7\,mm) using the Owen's Valley Radio Observatory's millimeter
array (OVRO) in C (baselines ranging from $7-22\,\mbox{k}\lambda$) and
L ($6-44\,\mbox{k}\lambda$) configurations during 18 tracks during the
years 2000-2002 (typical length of one track:~8 hours). To cover most
of the CO emission detected in our earlier observations using the IRAM
30\,m single dish telescope (Walter \& Heithausen 1999, Heithausen \&
Walter 2000) we observed 4 pointings (primary beam size of one OVRO
antenna: $\sim$1$'$).  The northern pointing (which was centred on the
maximum of the IRAM 30\,m observations) was observed for 10 tracks in
2000, the additional three pointings towards the south were added in
2002 (total: 8 tracks). Data were recorded using a correlator setup
resulting in velocity resolutions of 2.6 \kms\ (after online Hanning
smoothing) with a total bandwidth of $\sim150$ \kms\ (i.e. fully
covering the profiles detected in the single dish observations). Flux
calibration was determined by observing 1328+307 (3C286) and Neptune
for approximately 20 minutes during the observing runs.  These
calibrators and an additional noise source were used to derive the
complex bandpass corrections. The nearby calibrators B0716+714,
J1048+717 and J0841+708 were used as secondary amplitude and phase
calibrators. The data for each array were edited, calibrated and
mapped with the {\sc mma} and the {\sc miriad} packages.

The most northern pointing has a beamsize of 7.0$''\times$ 6.0$''$
(natural weighting) and an rms of $\sim$20\,mJy\,beam$^{-1}$ in a
2.6\,km\,s$^{-1}$ channel (conversion factor:
1\,K=0.45\,Jy\,beam$^{-1}$). The southern pointings have slightly
higher resolution (6.2$''\times$5.4$''$, due to the different
uv--coverage) but have significantly higher noise
($\sim$65\,mJy\,beam$^{-1}$, due to the shorter integration times).
The emission in individual channels is faint and all the detected
features are narrow (typically 3--5 channels); therefore it is not
possible to calculate integrated CO maps directly from the 60 channel
data cube. We thus blanked emission that was not present in
consecutive channels to create a map of the integrated CO emission.

\section{\ion{H}{2} Regions and Molecular Gas in the Tidal Arm}

Fig.~1 shows a B--band image of the region around NGC\,3077; the
contours show the distribution of neutral hydrogen (Walter et al.\ 
2002, here smoothed to 40$''$ resolution).  Fig.~2 is a blown--up
H$\alpha$ image of the region indicated by a box in Fig.~1 and shows
the newly detected \ion{H}{2} regions in the tidal arm near NGC\,3077.
In total, we catalogued 36 \ion{H}{2} regions -- all are associated
with the prominent tidal \ion{H}{1} arm near NGC\,3077.  For each
individual region we determined the coordinates, fluxes, luminosities,
diameters and \ion{H}{1} column densities (Tab.~1). In the following
we will adopt a distance of 3.6~Mpc to the the tidal arm (leading to a
scale of 17.5\,pc/[$''$]).  This corresponds to a distance modulus of
27.80~mag (values used in previous studies for M\,81, NGC\,3077 and
the surroundings: 27.76~mag: Sharina 1991, 27.93~mag: Sakai \& Madore
2001, 27.80~mag: Freedman \etal 1994).

\subsection{Global Comparison to Atomic and Molecular Gas}

The \ion{H}{2} regions are distributed over an impressive area of
$\sim4\times6$\,kpc$^{2}$=24\,kpc$^{2}$. Most of the \ion{H}{2}
regions are found where the \ion{H}{1} column density reaches values
$\>1.0\times10^{21}$\,cm$^{-2}$ (in a 13$''$ [$\sim$ 200\,pc] beam,
Tab.~1). Most of the regions are rather compact (Tab.~1), we derive an
average diameter for the regions (here defined as the FWHM of the
\ion{H}{2} regions, after deconvolving for the point spread function
of our H$\alpha$ observations, FWHM: 1.3$''$) of 1.9$''$ (2.3$''$
before the deconvolution). This corresponds to a linear size of
$\sim$35\,pc, i.e., similar to the radius of a Stromgen sphere of a
massive O star.

The thick lines in Fig.~2 represent the distribution of molecular gas
in the tidal features as traced by the CO(1$\to$0) emission observed
with the IRAM 30\,m single dish telescope (Heithausen \& Walter 2000).
The southern region coincides with the highest density of star
formation.  The higher star formation rate in the southern molecular
complex is also evident from the excitation conditions of the
molecular gas: the CO(2$\to$1)/CO(1$\to$0) line ratio is higher in the
south as compared to the regions in the north (Heithausen \& Walter
2000).

\subsection{Distribution of Molecular Gas at High Resolution}

Fig.~3 shows the distribution of the molecular gas as seen by the OVRO
interferometer (contours).  The circles indicate the primary beam
sizes of our four pointings.  Recall that the noise in the northern
pointing is a factor 3 lower than in the remaining pointings
(Sec.~2.2). In total we catalog six regions (A--F) - regions C, D and
F are marginal detections and need confirmation by follow-up
observations. The properties of these six regions as derived from the
OVRO observations (centre coordinates, fluxes, adopted masses and
systemic velocities) are summarized in Tab.~2. In Fig.~4 we compare
the OVRO (CO), IRAM 30\,m (CO) and VLA (HI) spectra for each of the
six regions. The OVRO spectra (thick full line) are plotted in units
of Jy\,beam$^{-1}$, as indicated on the left hand axis, the IRAM 30\,m
data (dashed line) are in units of T$_{\rm mb}$ (right hand axis). For
comparison, the VLA HI data are plotted as the thin dotted line.  From
this it is evident that the CO velocities are in good agreement with
the peaks of the \ion{H}{1} spectra at the same position. It is also
clear that regions C, D and F need follow--up confirmation.

Although the signal--to--noise of the OVRO observations is low, we can
still attempt to derive a virial mass estimate for the brightest two
regions (regions A and B). From the intergrated CO map we derive an
effective (deconvolved) radius $r$ of $\sim85$\,pc for both regions.
The line width $v$ for both regions is also similar (FWHM: $\sim$
7.5\,km\,s$^{-1}$). Using M$_{\rm vir}$=250$\times r$[pc]$\times
v$[km\,s$^{-1}$]$^2$ we thus derive approximate virial masses of $\sim
1\times10^6$\,\Msun\ for both regions (with an error of order
30--50\%). If we take the measured CO flux (S$_{\rm CO}$) of these two
regions and apply a Galactic conversion factor to convert CO
luminosities to molecular gas masses we derive H$_2$ masses of
1.3$\times10^6$\,\Msun\ and $0.75\times10^6$\,\Msun\ for regions A and
B, respectively (see caption of Tab.~2 for details). As these numbers
(virial mass compared to mass derived using a Galactic conversion
factor) are in agreement within the errors, this implies that the
CO--to--H$_2$ conversion factor in the tidal arm appears to be similar
to the Galactic value.  We note that a Galactic conversion factor was
also found to hold for the centre of NGC\,3077 (Meier et al.\ 2001,
Walter et al.\ 2002).  We thus derive a total molecular gas mass
recovered by the OVRO observations of 2.4$\times10^6$\,\Msun\ (using a
Galactic conversion factor for all regions, Tab.~2).

The difference in morphology between the CO map obtained at the IRAM
30\,m single dish telescope (Fig.~2) and the OVRO interferometer
(Fig.~3) is striking. It can be explained by the fact that our
interferometer observations are only sensitive to compact and clumpy
structures and will not recover more diffuse emission. In addition,
the 30\,m single dish map is not fully sampled (see Heithausen \&
Walter 2000 for details).  With our OVRO observations we only recover
about one sixth of the total flux detected by the earlier 30\,m
observations (there we find a total flux of of 22 K\,km\,s$^{-1}$ [100
Jy\,km\,s$^{-1}$] in the area covered by OVRO\footnote{Note that the
  molecular gas masses quoted in Heithausen \& Walter 2000 were
  derived using a conversion factor that was $\sim$3 times higher than
  the Galactic factor used here.}) which is entirely consistent with
the picture that most of the molecular gas is more extended.  Only one
molecular cloud complex is associated with a bright \ion{H}{2} region
(complex A, region 17).  Regions C and D are also associated with
faint H$\alpha$ emission, however most of the other prominent
\ion{H}{2} regions, in particular in the southern pointing, do not
show corresponding CO emission down to our sensitivity limit. One
bright CO complex (B) has no associated star formation. It is likely
that new star formation will commence in this region in the future.
Given the (likely) low metallicity in the tidal arm, we don not expect
extinction to play a major role in these regions. Observations of the
dust phase (e.g. using Spitzer) would be needed to check if highly
extincted, embedded star formation may be present in some areas of the
tidal feature.

\subsection{\ion{H}{2} Luminosity Function}

We now investigate how the properties of the 36 \ion{H}{2} regions in
the tidal feature compare to \ion{H}{2} regions found in other dwarf
galaxies in the same group.  The luminosity function of the \ion{H}{2}
regions in the tidal arm is presented in Fig.~5 (thick histogram).
Also plotted in this figure are the \ion{H}{2} luminosity functions
derived for other M\,81 group dwarf galaxies by Miller \& Hodge
(1994).  Note that the decrease in the luminosity function towards low
luminosities is due to the sensitivity of the observations (down to a
completeness limit of log(L[erg\,s$^{-1}$])$\sim$36.4) for both our
observations and the observations presented in Miller \& Hodge (1994).
At the bright end of the luminosity function, values for \ion{H}{2}
regions in the tidal feature do not reach values as high as found in
big spiral galaxies or more massive dwarf galaxies (e.g., IC\,2574).
However, keeping in mind the low number statistics, the shape of the
\ion{H}{2} region luminosity function is similar to what is found in
other low--mass dwarfs such as Holmberg~I (Fig.~5).

Even for our brightest \ion{H}{2} regions the H$\alpha$ luminosity can
in principle be emitted by a single massive star (L$_{\rm H\alpha,
  O7}\approx5\times10^{36}$\,erg\,s$^{-1}$, L$_{\rm H\alpha,
  O5}\approx5\times10^{37}$\,erg\,s$^{-1}$, Devereux et al.\ 1997).
The total H$\alpha$ and [NII] flux (luminosity) for all the \ion{H}{2}
regions is:
F(H$\alpha)$=1.88$\times10^{-13}$\,erg\,s$^{-1}$\,cm$^{-2}$
(L(H$\alpha)$=2.91$\times10^{38}$\,erg\,s$^{-1}$). This corresponds to
a total star formation rate (SFR) in the entire tidal arm of SFR $
\approx 2.3\times10^{-3}$M$_\odot$ yr$^{-1}$, where SFR(0.1-100\Msun\ 
stars) = L(H$\alpha$)/1.26$\times10^{41}$\,erg\,s$^{-1}$ [M$_\odot$
yr$^{-1}$] (Kennicutt 1998). The ionizing stars are part of stellar
clusters, as revealed by deep ACS imaging obtained with the HST
(proposal ID 9381, these data will be presented elsewhere).

\section{Discussion and Summary}

We present the discovery of wide-spread, low-level star formation in
the tidal feature near NGC\,3077 (total of 36 \ion{H}{2} regions).
The \ion{H}{1} column densities exceed $1\times10^{21}$\,cm$^{-2}$
over the $\sim$24\,kpc$^2$ region where the \ion{H}{2}\ regions are
found.  The total star formation rate in the tidal arm is
SFR=$2.3\times10^{-3}$\,M$_\odot$ yr$^{-1}$.  These properties place
the complexes amongst the faintest star forming regions detected in
tidal arms around galaxies so far (cf. Ryan-Weber et al.\ 2004).  The
presence of a huge reservoir of molecular gas ($\sim 10^7$\,\Msun,
Heithausen \& Walter 2000, using the Galactic conversion factor) as
well as atomic gas (M$_{\rm HI}\sim3\times10^8$\,\Msun\ in the region
shown in Fig.~2) suggests that star formation may continue at the
current rate for a Hubble time. If some 10 percent of the gas were
transformed into stars, the total luminosity of the tidal dwarf will
eventually be of order $5\times10^{7}$\,M$_{\rm \odot}$ (here we
simplisticly adopt a stellar mass--to--light ration of 1\,M$_{\rm
  \odot}$/L$_{\rm \odot}$), similar to other faint dwarf galaxies in
the M\,81 group.

It is interesting to note that no clear correlation exists between the
sites of on--going star formation and gas column densities (both
molecular and atomic): \ion{H}{2} regions are not always found where
the \ion{H}{1} or molecular column density is highest (e.g., Fig.~3).
However, star formation is only seen where the \ion{H}{1} column
density reaches values larger than $\sim 1.0\times10^{21}$\,cm$^{-2}$
(Tab.~1) -- this value is consistent with what is typically found in
(dwarf) galaxies (e.g., Skillman 1987, Walter \& Brinks 1999, Martin
\& Kennicutt 2001).  Also, the derived \ion{H}{2} luminosity function
is, within the errors, similar to other low-mass dwarf galaxies in the
same group (such as Holmberg~I). No \ion{H}{2} regions are found with
log(L[erg\,s$^{-1}$])$>$37.4 -- this is more than an order of
magnitude fainter than what is found in more massive dwarf galaxies in
the same group (e.g. IC\,2574) and bigger spiral galaxies (e.g., in
M51 luminosities reach values of log(L)[erg\,s$^{-1}$]=39, e.g., Rand
1993). None of the \ion{H}{2} regions are coincident with the X--ray
sources found in the same region of the sky (Ott, Martin \& Walter
2003).

It is difficult to estimate when star formation started in the tidal
arm. From the fact that we see H$\alpha$ emission today it is clear
that star formation is occuring well after the creation of the
structure some $\sim3\times 10^8$\,yr ago (i.e., `in situ' formation).
If we assume a constant star formation rate over the last
$3\times10^8$\,yr we estimate a total mass of newly formed stars of
$\sim7\times10^{5}$\,M$_{\odot}$.  This star formation is enriching
the ISM of the tidal tail with heavy elements: Taking our derived
assembled stellar mass since the creation of the feature, and a yield
of 2\%, we derive a metal production of 1.4$\times$10$^4$\,\Msun\ over
the last $3\times10^8$\,yr. We divide this number by the total amount
of gas within the $\sim$4$\times$6\,kpc region enclosed by the
10$^{21}$\,cm$^{-2}$ contour (3$\times10^8$\,\Msun) to get a
metallicity of 4.7$\times10^{-5}$, or roughly Z$\sim$0.002 Solar, well
below the values measured for the gas phase in the most metal--poor
dwarfs.  We note that the total metallicity in the tidal feature is
likely higher as models for the tidal arm show that it was stripped
from the outskirts of NGC\,3077 (Thomasson \& Donner 1993, Brouillet
\etal 1991, Yun \etal 1993).  Measurements of the gas phase
metallicity in the tidal feature would thus clearly help to shed light
on its origin and history.  We conclude that wide--spread low--level
star formation may be a common phenomenon in tidal \ion{H}{1} tails
with column densities that exceed $1\times10^{21}$\,cm$^{-2}$, leading
to chemical enrichment in these tails.

This type of low-level star formation in tidal debris should be
detectable, if it is common, as intervening absorption in quasar
spectra.  The tidal system discussed here would be classified as a
damped Lyman--alpha absorber (DLA) with a cross section of roughly 30
kpc$^2$. A tidal origin for (at least some) DLAs would naturally
explain their lack of enhanced [$\alpha/$Fe] abundance ratios, which
suggest that star formation proceeded relatively slowly (Pettini et
al.\ 1999).  Direct observations of DLAs at low redshift are
consistent in that they indicate associations with a wide variety of
galaxy types and relatively small impact parameters (of order 10\,kpc,
e.g., Kanekar \& Chengalur 2003).  For comparison, the tidal system
discussed here has a mean projected distance of $\sim$5\,kpc from NGC
3077 ($\sim$50\,kpc distance from M\,81). This interesting object
should be further explored via metallicity measurements and
simulations that would establish the fate of the metals formed in this
system.

\acknowledgements CLM thanks the David and Lucile Packard Foundation
and the Alfred P.  Sloan Foudation for supporting this work. We thank
Bryan Miller for sharing his \ion{H}{2} luminosity functions with us
and the anonymous referee for providing useful comments. Research with
the Owens Valley Radio Telescope, operated by Caltech, has been
supported by NSF grant AST96--13717. Support of this work was also
provided by a grant from the K.T. and E.L. Norris Foundation.

\begin{figure}
\begin{center}
\epsscale{0.5}
\plotone{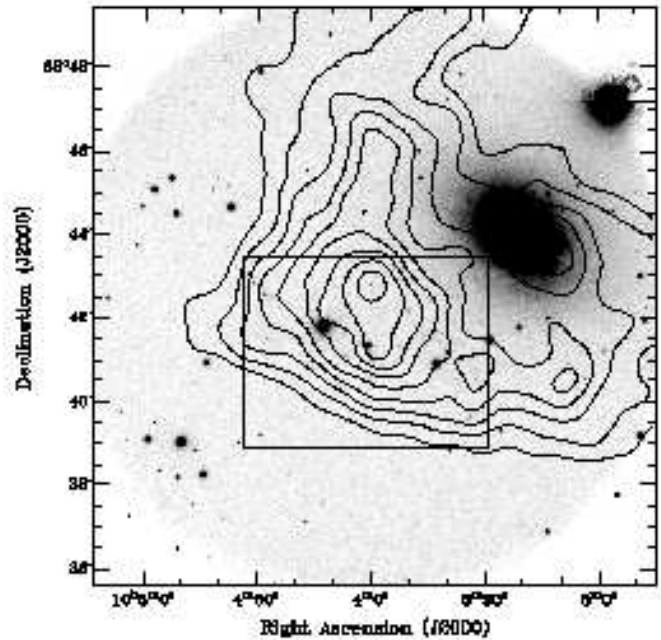}
\caption{B--band image of
  NGC\,3077 and its surrounding. Overlaid are the \ion{H}{1} contours
  which show the orientation of the tidal arms (smoothed to 40$''$,
  the 10$^{21}$\,cm$^{-2}$ column is the third lowest contour -- the
  peak column density at 40$''$ resolution in the tidal feature is
  $\sim$2.5$\times$10$^{21}$\,cm$^{-2}$, Walter et al.\ 2002). The box
  marks the region where \ion{H}{2} regions in the tidal arm have been
  detected (see Fig.~2).}
\end{center}
\end{figure}

\begin{figure}
\begin{center}
  \epsscale{1}
  \plotone{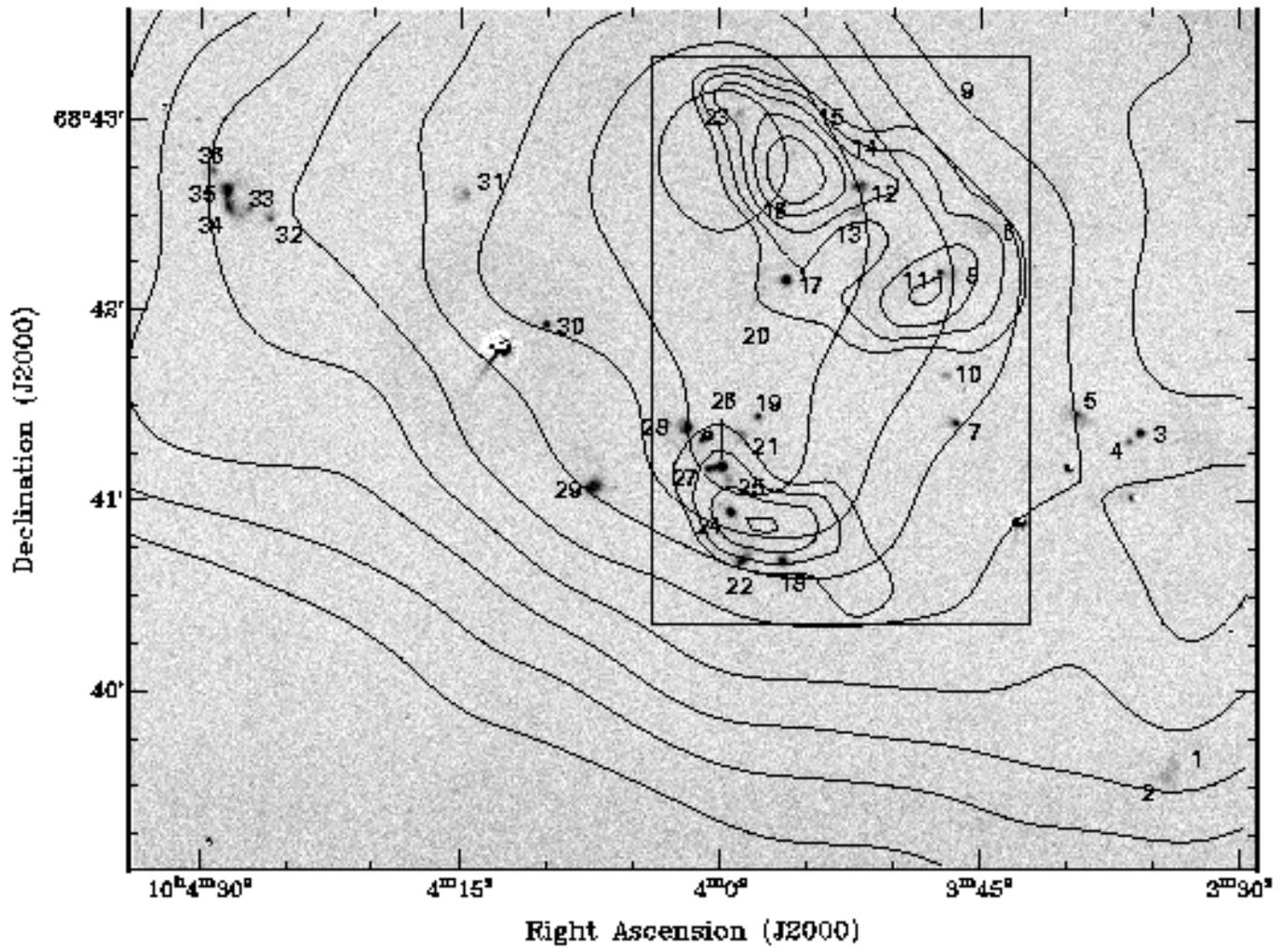} \caption{H$\alpha$ image
    of the tidal arm near NGC\,3077 (the region is indicated by the
    box in Fig.~1). In total, 36 \ion{H}{2} regions were catalogued.
    The thin contours again indicate the \ion{H}{1} emission (smoothed
    to 40$''$ resolution). The thick contours represent the CO(1--0)
    emission in the the tidal arm as observed with the IRAM 30\,m
    single dish telescope (only the box has been searched for CO,
    Walter \& Heithausen 1999, Heithausen \& Walter 2000).}
\end{center}
\end{figure}

\begin{figure}
\begin{center}
\epsscale{0.5}
  \plotone{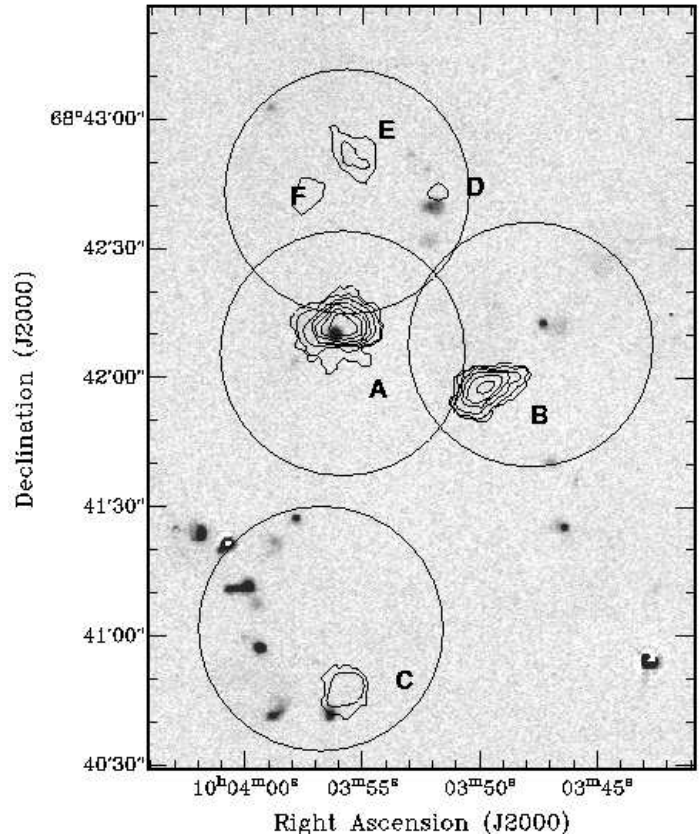} \caption{Close--up view
    of the \ion{H}{2} regions (region is shown as a box in Fig.~2).
    Here the contours represent the CO(1--0) emission as seen by our
    interferometric observations using OVRO (derived from the blanked
    data cube). CO contours are shown at 0.2, 0.4, 0.8, 1.2, 1.6, 2.0,
    2.4 Jy\,beam$^{-1}$\,km\,s$^{-1}$.  The beamsize in regions A, B
    \& C (7.0$''\times6.0''$) is slightly different from those of D, E
    \& F (6.2$''\times$5.4$''$, see Sec.~2.2). The circles indicate
    the primary beam sizes of the four OVRO pointings. Regions C, D
    and F are only marginal detections and need follow-up
    confirmation.}
\end{center}
\end{figure}

\begin{figure}
\begin{center}
  \epsscale{1.0}
  \plotone{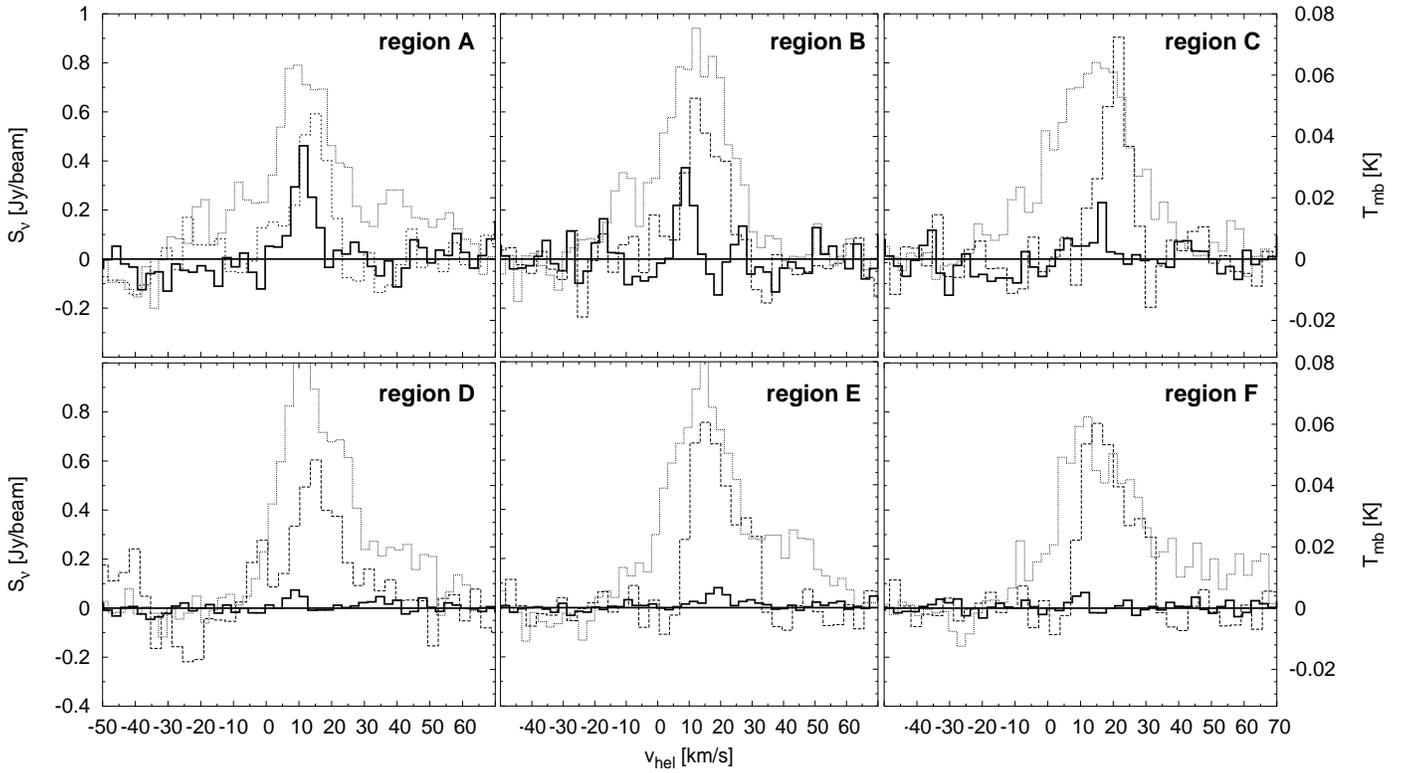} \caption{Spectra for the six regions labeled in
    Fig.~3. The OVRO spectra (thick full line) are plotted in units of
    Jy\,beam$^{-1}$, as indicated on the left hand axis, the IRAM
    30\,m data (dashed line) are in units of T$_{\rm mb}$ (right hand
    axis). For comparison, the VLA HI data are plotted as the thin
    dotted line (arbitrary units). The noise in the OVRO observations
    of regions D, E and F is signifcantly less than in the other three
    regions as more time was spent on the most northern OVRO pointing
    (see text). The physical parameters derived from the OVRO
    observations are summarized in Tab.~2.}
\end{center}
\end{figure}

\begin{figure}
\begin{center}
  \epsscale{0.4}
  \plotone{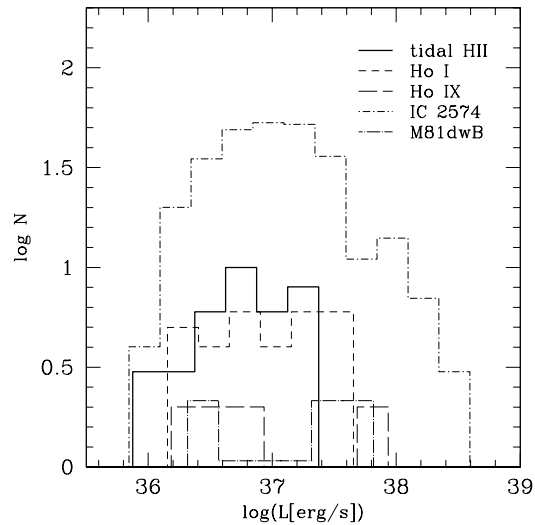} \caption{Luminosity
    function of the \ion{H}{2} regions in the tidal arm compared to
    other dwarf galaxies in the same group (Miller \& Hodge 1994,
    corrected for our adopted distance). The histograms have been
    offset slightly for better readability. Our \ion{H}{2} region
    catalog is not complete below \ion{H}{2} luminosities fainter than
    $\sim$ 36.4\,erg\,s$^{-1}$ (similar to Miller \& Hodge 1994).}
\end{center}
\end{figure}

\clearpage

\begin{table}
\scriptsize
\begin{center}
\caption{Properties of the \ion{H}{2} regions}
\begin{tabular}{rllcccc}
\tableline \tableline
Nr.&    RA (2000)   &   DEC (2000)  & F(H$\alpha)^{a,b}$   & log(L(H$\alpha)$)  &  D$^c$ &  $\sigma_{\rm HI}^d$  \nl
   & hh~mm~ss       &   $\circ$~$'$~$''$  &  10$^{-15}$erg\,s$^{-1}$\,cm$^{-2}$  & erg\, s$^{-1}$ & $''$ & 10$^{21}$\,cm$^{-2}$ \nl
\tableline                                                 
1 &     10 03 33.73 &   68 39 37.9  &  1.98  &      36.5   &  ---   &   0.9  \nl
2 &     10 03 34.18 &   68 39 33.6  &  3.47  &      36.7   &  ---   &   0.7  \nl
3 &     10 03 35.68 &   68 42 21.9  &  7.63  &      37.1   &  1.6   &   0.9  \nl
4 &     10 03 36.35 &   68 41 19.5  &  2.00  &      36.5   &  1.9   &   1.0  \nl
5 &     10 03 39.31 &   68 41 28.0  &  8.92  &      37.1   &  4.8   &   1.2  \nl
6 &     10 03 44.56 &   68 42 27.4  &  6.52  &      37.0   &  ---   &   1.5  \nl
7 &     10 03 46.29 &   68 41 25.0  &  4.13  &      36.8   &  1.8   &   1.5  \nl
8 &     10 03 46.51 &   68 42 12.1  &  2.06  &      36.5   &  ---   &   1.8  \nl 
9 &     10 03 46.68 &   68 43 09.7  &  0.49  &      35.9   &  1.3   &   0.9  \nl
10 &    10 03 46.85 &   68 41 40.2  &  1.37  &      36.3   &  2.1   &   1.5  \nl
11 &    10 03 47.24 &   68 42 12.5  &  3.45  &      36.7   &  1.8   &   1.8  \nl
12 &    10 03 51.88 &   68 42 39.8  &  6.87  &      37.0   &  3.2   &   2.1  \nl
13 &    10 03 52.10 &   68 42 31.9  &  2.25  &      36.5   &  2.9   &   2.0  \nl
14 &    10 03 52.16 &   68 42 48.6  &  0.56  &      35.9   &  1.3   &   2.0  \nl
15 &    10 03 52.89 &   68 42 52.0  &  0.79  &      36.1   &  1.7   &   1.9  \nl
16 &    10 03 55.74 &   68 42 50.4  &  1.00  &      36.2   &  ---   &   1.7  \nl
17 &    10 03 56.18 &   68 42 10.3  &  11.9  &      37.3   &  1.7   &   1.6  \nl
18 &    10 03 56.38 &   68 40 42.0  &  6.23  &      37.0   &  2.3   &   1.6  \nl
19 &    10 03 57.79 &   68 41 27.4  &  3.13  &      37.7   &  1.5   &   2.3  \nl
20 &    10 03 57.85 &   68 42 03.9  &  0.97  &      36.2   &  2.0   &   1.9  \nl
21 &    10 03 58.68 &   68 41 21.9  &  3.73  &      36.8   &  ---   &   2.5  \nl
22 &    10 03 58.72 &   68 40 41.7  &  9.55  &      37.2   &  2.7   &   1.3  \nl
23 &    10 03 58.93 &   68 43 02.6  &  1.93  &      36.5   &  2.5   &   2.5  \nl
24 &    10 03 59.34 &   68 40 57.2  &  10.9  &      37.2   &  1.8   &   1.8  \nl
25 &    10 03 59.46 &   68 41 07.0  &  2.61  &      36.6   &  ---   &   2.5  \nl
26 &    10 03 59.85 &   68 41 11.2  &  11.8  &      37.3   &  2.1   &   2.7  \nl
27 &    10 04 00.63 &   68 41 10.6  &  5.53  &      36.9   &  2.1   &   2.5  \nl
28 &    10 04 01.92 &   68 41 24.0  &  0.12  &      37.3   &  2.5   &   1.8  \nl
29 &    10 04 07.33 &   68 41 05.1  &  0.15  &      37.4   &  4.1   &   1.5  \nl
30 &    10 04 10.02 &   68 41 56.5  &  4.28  &      36.8   &  1.5   &   1.7  \nl
31 &    10 04 14.73 &   68 42 37.8  &  4.13  &      36.8   &  3.2   &   1.4  \nl
32 &    10 04 26.02 &   68 42 29.7  &  3.10  &      36.7   &  2.4   &   1.2  \nl
33 &    10 04 27.25 &   68 42 32.4  &  3.46  &      36.7   &  2.4   &   1.0  \nl
34 &    10 04 28.37 &   68 42 33.0  &  8.57  &      37.1   &  --    &   0.9  \nl
35 &    10 04 28.54 &   68 42 38.5  &  0.12  &      37.3   &  3.3   &   1.0  \nl
36 &    10 04 29.32 &   68 42 44.6  &  3.35  &      36.7   &  2.4   &   0.9  \nl
\tableline \tableline
\tablecomments{\\$^a$ The H$\alpha$ fluxes include emission from [NII] $\lambda \lambda 6548,6583$ (see discussion in text).  Correction for Galactic extinction would raise the H$\alpha$ fluxes
by 11\%.\\
$^b$ Typical errors (1 $\sigma$) are 0.04$\times10^{-15}$ erg\,s$^{-1}$\,cm$^{-2}$. \\
$^c$ Diameters are here defined as the effective FWHM of the emission; no entry indicates diffuse emission; values given here are not deconvolved for seeing (1.3$''$, see Sec.~3.1), errors are of order 0.5$''$.\\
$^d$ \ion{H}{1} column densities are derived from the 13$''$ map presented in Walter et al.\ (2002). Systematic uncertainties of these values are of order 10\%.
}
\end{tabular}
\end{center}
\end{table}

\normalsize

\begin{table}
\caption{Properties of compact molecular cloud complexes}
\begin{center}
\begin{tabular}{lllccc}
\tableline \tableline
region & RA         & DEC        & flux            & H$_2$ mass$^b$         & v$_{\rm hel}^{c,d}$  \nl   
       & (J2000.0)  & (J2000.0)  & Jy km\,s$^{-1}$ & 10$^4$ M$_\odot$ & km\,s$^{-1}$ \nl
\tableline                                                 
A      & 10 03 55.8 & 68 42 11.8 &  8.2            &  130             & 12.7 \nl
B      & 10 03 49.7 & 68 41 57.8 &  4.7            &   75             & 10.1 \nl
{\em C}$^a$  & 10 03 55.7 & 68 40 47.5 &  1.2            &   19             & 17.9 \nl
{\em D}$^a$  & 10 03 51.8 & 68 42 42.9 &  0.15           &   2.4            & 10.1 \nl
E      & 10 03 55.5 & 68 42 50.8 &  0.90           &   14             & 20.5 \nl
{\em F}$^a$  & 10 03 57.4 & 68 42 42.5 &  0.35           &   5.6            & 12.7 \nl
\tableline \tableline
\tablecomments{\\$^a$ Marginal detections. Note that the sensitivity of the observations for regions D, E and F is three times higher than for the other regions (see text for details).\\
  $^b$ Molecular masses are derived using
  M=1.23$\times$10$^4\times$(3.6)$^2\times$ S$_{\rm CO}$\,M$_{\odot}$. This
  assumes a Galactic CO-to-H$ _2$ conversion factor of $2.3
  \times10^{20}$\,cm$^{-2}$\,(K\,km\,s$^{-1}$)$^{-1}$, Strong et al.\ 
  (1988).\\
  $^c$ Channel with strongest emission.\\
  $^d$ typical linewidths: 3 channels, i.e. $\sim$7.5\,km\,s$^{-1}$.
}
\end{tabular}
\end{center}
\end{table}

\end{document}